# Privacy Mining from IoT-based Smart Homes

Ming-Chang Lee
Department of Communication Systems, Simula Research Laboratory
Martin Linges vei 25, Fornebu, 1364, Norway
mclee@simula.no

Jia-Chun Lin and Olaf Owe
Department of Informatics, University of Oslo
Gaustadallèen 23 B, Oslo, N-0373, Norway
{kellylin, olaf}@ifi.uio.no

September 10, 2018

# Privacy Mining from IoT-based Smart Homes

Ming-Chang Lee[1], Jia-Chun Lin[2], Olaf Owe[2]

[1] Department of Communication Systems, Simula Research Laboratory
Martin Linges vei 25, Fornebu, 1364, Norway
mclee@simula.no

[2] Department of Informatics, University of Oslo
Gaustadallèen 23 B, Oslo, N-0373, Norway
{kellylin, olaf}@ifi.uio.no

**Abstract.** Recently, a wide range of smart devices are deployed in a variety of environments to improve the quality of human life. One of the important IoT-based applications is smart homes for healthcare, especially for elders. IoT-based smart homes enable elders' health to be properly monitored and taken care of. However, elders' privacy might be disclosed from smart homes due to non-fully protected network communication or other reasons. To demonstrate how serious this issue is, we introduce in this paper a Privacy Mining Approach (PMA) to mine privacy from smart homes by conducting a series of deductions and analyses on sensor datasets generated by smart homes. The experimental results demonstrate that PMA is able to deduce a global sensor topology for a smart home and disclose elders' privacy in terms of their house layouts.

## 1   Introduction

The Internet of Things (IoT) is a technology paradigm envisioned as a global network of machines and devices capable of connecting and interacting with each other [1]. It is expected that we will have 24 billion IoT devices by 2020 [2]. IoT enables different smart environments with various smart devices [3][4] such as smart homes, smart buildings, smart communities, smart cities, and smart grids. IoT also enables numerous applications, including healthcare [4][5][6], home automation [7], security [8][9][10], and surveillance [11][12][13][14]. Among these applications, healthcare attracts a lot of attention in recent years [4][15] due to the growth of the elderly population. Using smart home technology, elders' health can be taken care of, and healthcare costs can be dramatically reduced.

General speaking, a smart home for elder healthcare consists of motion sensors, door sensors, and other sensors, which all connect to a smart hub via Z-Wave or ZigBee. The hub also connects to a router via WiFi or Ethernet so as to communicate with a cloud server on the Internet. Data sensed by all different sensors is transmitted to the cloud server and analyzed by the cloud server. In such network environments, the sensor data of smart homes might be eavesdropped by malicious people using sophisticated tools [16][17][18][19]. The sensor data might also be completely exposed to malicious people who are able to access the cloud server.

In this paper, we are interested in knowing what privacy information we can mine from a smart home giving that we can access a set of sensor dataset that only provides *limited* information about a smart home. By *limited* we mean that each dataset record consists of exactly a timestamp and a fingerprint that refer to a particular sensor triggered/activated at a given point in time. Even when all the transmissions are encrypted, one may obtain this dataset using the Fingerprint and Timing-based Snooping (FATS) attack, which is an attack using information revealed by a cryptographic system rather than using the ciphertext to infer either the cryptographic keys or the original data [33]. To mine privacy from the sensor dataset, we propose a Privacy Mining Approach (PMA for short) based on a data mining technique and a series of deductions. More specifically, PMA defines how to identify elders' movement activities from the sensor dataset, and then deduces a global sensor topology based on all identified activities. Finally, PMA derives the locations where the sensors are deployed (e.g., bedrooms) by using Association Rule Learning [20].

To demonstrate the deduction performance of PMA, we applied PMA on a sensor dataset generated by a real smart home. The experimental results show that PMA is able to deduce a global sensor topology that corresponds to the real one for the smart home, and is able to infer most sensors in the bedroom and kitchen/dining room of the smart home with a high accuracy rate.

The rest of this paper is organized as follows. Section 2 describes the related work. Section 3 presents the design of PMA. In Section 4, extensive experiments are conducted and experimental results are discussed. Section 5 gives a conclusion and suggests future work of this paper.

## 2 Related Work

To recognize common human activities from sensor reading, different activity recognition methods have been proposed. Typically, activity recognition can be either vision-based or sensor-based [21]. The former uses visual sensing facilities, such as video cameras, to monitor people's behavior and environmental changes. However, it is well known that this approach suffers from privacy and ethics issues [6][21][22]. The latter employs embedded sensors (such as motion sensors and door sensors) or wearable sensors (such as RFID tags attached to homeowners). The focus in this paper is to mine privacy from sensor data sent from embedded sensors since we are interested in understanding how elders move from sensor to sensor, which allows us to deduce the global sensor layout of the corresponding smart homes.

Activity recognition can be classified into two types: supervised learning and unsupervised learning. In a supervised activity recognition method [23][24][25], each sensor event has its activity label, such as personal hygiene, enter home, bathing, bed to toilet, etc. For instances, Zhao et al. [26] introduce a CRFs-based classifier to recognize human activities. Inomata et al. [27] utilized Dynamic Bayesian Network framework to recognize activities from interaction data collected by a RFID tag system. Lu et al. [6] proposed a method for extracting latent features from sensor data by using Beta Process Hidden Markov Model (BP-HMM). However, it is well known that all the supervised methods suffer from the problem of manually labeling for all activities in training phase [15].

To address this problem, unsupervised methods [5][15][28][29][30] have been proposed to automatically recognize human activities in smart homes. Rashidi et al. [5] presented an unsupervised method for discovering and tracking activities that homeowners normally perform in smart homes. Gu et al. [29] proposed a fingerprint-based algorithm to recognize activities such that it can mine large number of activity models on the web without manual labeling. Rashidi and Cook [30] introduced a stream mining method for automatically discovering human activity patterns over time from streamed sensor data.

Different from all of the above approaches, in this paper, we are not interested in what exactly each activity does (i.e., the label of each activity) since our goal is not to recognize elders' different activities. Instead, we attempt to identify elders' movements from a sensor dataset because it reveals how elders move in their smart homes, enabling us to reason about the relationship between sensors and deduce a global sensor topology.

Srinivasan et al. [33] presented a multi-tier FATS algorithm to infer information about a home and its residents based on a dataset eavesdropped by the FATS attack. The dataset consists of just the timestamps and fingerprints of radio transmissions. The authors claimed that they are able to identify the function of each room by using a training data from other houses in which each room is labeled as a bedroom, kitchen, bathroom, or living room and then using a number of features for every room cluster to infer an unknown room in the target house. However, this approach is contradictory to their employment of the FATS attack because a dataset eavesdropped by the FATS attack does not reveal any label information for each room. It is unclear how to get the training data with room labels without human intervention.

In contrast to the multi-tier FATS algorithm, in this paper, we do not require any training data or any information with room labels to reason about the function of a room. Instead, we identify all indoor activities to reason about a global sensor topology, and then mine bedroom sensors and living room/dining room sensors based on the derived global sensor topology and common human habits.

## 3  The Design of PMA

In this section, we explain how PMA mines a global sensor topology for a smart home and reasons about sensor locations by assuming that a set of sensor dataset generated by a smart home is available. As mentioned earlier, each dataset record indicates a timestamp and a fingerprint that refers to a particular sensor triggered/activated at that moment. For readability, in this paper, each unique fingerprint is replaced by a unique sensor ID before employing PMA.

### 3.1  Sensor Topology Deduction

In this paper, an elder's movement activity in a smart home is called an *indoor activity*. This movement lasts for a while and triggers a sequence of sensors. The rule to identify an indoor activity is defined as "*A sequence of sensors are triggered during a time period of at least $x$ seconds, and the interval between any two adjacent sensors is at most $y$ seconds where $x > y$*". For example, if $x = 40$ sec and $y = 10$ sec, the first six log records shown in Fig. 1 will be identified by PMA as an indoor activity because their duration is longer than 40 sec, and the interval periods between any two adjacent records are not longer than 10 seconds. However, the last record will not be considered as a part of the same activity because the interval period between the record and the previous record is longer than 10 seconds. Based on the above rule, PMA identifies all indoor activities from the sensor dataset.

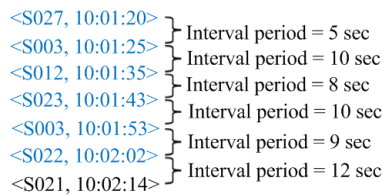

**Fig. 1.** An example of indoor activities (i.e., the first six records).

To deduce a global sensor topology, PMA first translates each indoor activity into a set of directed edges. For example, the indoor activity shown in Fig. 1 can be translated into the following set of directed edges:

$$S027 \rightarrow S003, S003 \rightarrow S012, S012 \rightarrow S023, S023 \rightarrow S003, S003 \rightarrow S022$$

This set shows that the elder moves from sensor to sensor when performing the indoor activity. After translating every identified indoor activity into a corresponding directed-edge set, PMA accumulates a confidence value $i$ for each directed edge $A \rightarrow B$. Note that $A$ and $B$ are two different sensors in the smart home, and $i$ the total number of occurrences of $A \rightarrow B$ appearing in all the directed-edge sets. If $i$ is large, we can claim that the elder can move directly from sensor $A$ to sensor $B$ with high confidence. Based on the following two rules, a global sensor topology can be therefore constructed:

*Rule 1*: If both $A \rightarrow B$ and $B \rightarrow A$ have a confidence value larger than a predefined threshold $\alpha$, PMA is confident that the elders are *able* to directly move from $A$ to $B$ and vice versa. Therefore, PMA adds a bidirectional solid edge between $A$ and $B$. Note that $\alpha = \left\lfloor \frac{\beta}{\gamma} \right\rfloor$ where $\beta$ is the summation of the confidence values of all the directed edges, and $\gamma$ is the total number of all the directed edges. In other words, $\alpha$ is the average confidence value of all the directed edges.

*Rule 2*: If only $A \rightarrow B$ or only $B \rightarrow A$ has a confidence value larger than $\alpha$, PMA presumes that it is *possible* for the elder to directly move from $A$ to $B$ and vice versa. In this case, PMA adds a bidirectional dash edge between $A$ and $B$.

Apparently, Rule 2 appears less strict than Rule 1, but it enables PMA to derive a global view of the sensors since most sensors could thereby be connected together.

## 3.2 Sensor Location Reasoning

PMA employs Association Rule Learning (ARL) [20] to find sensor locations. ARL is a rule-based machine learning method designed to find groups of items that are commonly found together in a dataset. In the ARL method, the rule $X \Rightarrow Y$ holds with minSupport $s$ if at least $s * 100\%$ of the records in a dataset contains both $X$ and $Y$ where $0 < s < 1$. In other words, both $X$ and $Y$ are present together in at least $s * 100\%$ of the records in the dataset.

Fig. 2 illustrates the algorithm of reasoning about sensor locations. Let $w$ be the total number of days in the sensor dataset available to PMA. Note that PMA focuses on reasoning about smart homes with bedrooms and a kitchen/dining room. Reasoning about other rooms will be future work. In addition, we do not differentiate kitchen and dining room because in many houses they are in the same place.

As shown from line 1 to line 17, PMA first deduces bedroom sensors by reasonably assuming that most people stay in their bedrooms between 2 am and 6 am, implying that the sensors deployed in bedrooms are very likely to be triggered during this time period. For each indoor activity occurring between 2 am and 6 am of these $w$ days, PMA translates it into a sensor-ID list by extracting all distinct sensor IDs from the activity. For instance, the sensor-ID list of the indoor activity shown in Fig. 1 is {S003, S012, S022, S023, S027}.

```
The sensor-location deduction algorithm
Input: All indoor activities of a smart home in w days and global sensor topology
Output: Bedroom sensors and kitchen/dining room sensors.
Procedure:
1:   Let A be an empty set;
2:   for each indoor activity that occurred between 2 am and 6 am of the w days{
3:       Translate the activity into a sensor-ID list;
4:       Put the sensor-ID list into A;}
5:   Apply ARL with minSupport = 0.5 on A;
6:   Let i = 1;
7:   do {
8:       Choose a sensor set from the outcome of ARL if this set is the largest one and its
9:       occurrence is the most frequent compared with all other sets of the same size;
10:      Let this set be B and output B to be sensors in bedroom i;
11:      while a sensor in the topology has bidirectional edges with at least a half of B{
12:          Consider this sensor to be in bedroom i;
13:          Put the sensor into B;}
14:      Discard any sensor-ID list that contains any sensor of B from A;
15:      Apply ARL with minSupport = 0.5 on A again;
16:      i = i + 1;
17:  } while ARL outputs at least one sensor set & the size of the set is larger than 1;
18:  Let C be an empty set;
19:  for each indoor activity that occurred between 6 pm and 7 pm of the w days{
20:      Translate the activity into a sensor-ID list;
21:      Put the sensor-ID list into C;}
22:  Apply ARL with minSupport = 0.5 on C;
23:  Choose a sensor set from the outcome of ARL if this set is the largest one and its
24:  occurrence is the most frequent compared with all other sets of the same size;
25:  Let this set be K and output K to be sensors in kitchen/dining room;
26:  while a sensor in the topology has bidirectional edges with at least a half of K{
27:      Also consider this sensor to be a sensor in the kitchen/dining room;
28:      Put the sensor into K;}
```

**Fig. 2.** The sensor-location deduction algorithm.

Let $A$ be a set of sensor-ID lists that associates with all indoor activities occurring between 2 am and 6 am of the $w$ days. PMA then applies ARL with minSupport of 0.5 on $A$ to find out all possible sets of sensors that satisfy the minSupport, i.e., the majority. PMA chooses a set that is the largest one, and the occurrence of this set is the most frequent one as compared with all other sets of the same size (see lines 8 and 9). This set of sensors (denoted by $B$) will be deduced to be sensors in the same bedroom.

After that, PMA attempts to deduce more sensors in the same bedroom by using the global sensor topology. Based on the majority rule, if any sensor in the global sensor topology has bidirectional edges with at least a half of $B$, this sensor will be considered to be in the same bedroom. Furthermore, it is possible that a smart home has more than one bedroom. By removing all sensor-ID lists that contain any known bedroom sensor from $A$ and repeating the above steps, PMA is able to find sensors in other bedrooms.

PMA continues by deducing sensors in the kitchen/dining room (see lines 18 to 28) under an assumption that most people stay in their kitchens/dining rooms between 6 pm and 7 pm. This time period is typically time for dinner. Based on this assumption, PMA translates each indoor activity occurring within this time period of the $w$ days into a sensor-ID list and repeats the same procedure (i.e., using ARL and the global sensor topology) to deduce all possible sensors in the kitchen/dining room.

## 4 Experimental Results

To evaluate the ability of PMA in mining privacy from smart homes, we chose a real smart home provided by the WSU CASAS smart home project [31] to be our use case. The chosen smart home is named "Milan", and it has only one floor. The residents in this house are an old woman and a dog. Milan has a sensor dataset recording the time at which each sensor is triggered. The complete dataset contains 433,665 sensor records generated by 7 area motion sensors, 21 motion sensors, 3 door closure sensors, and 2 temperature sensors. In the following experiments, we ignored the sensor data generated by temperature sensors since this data is irrelevant, and we focused on applying PMA on only 6-days sensor data (from Oct. 16[th], 2009 to Oct. 21[th], 2009), i.e., $w = 6$. Fig. 3 illustrates the layout and sensor deployment of Milan. Our goal is to see if PMA is able to deduce the global sensor topology of Milan and infer sensors deployed in the bedrooms and kitchen/dining room without knowing the layout given in Fig. 3. In this experiment, PMA identified all indoor activities by setting $x = 40$ sec and $y = 10$ sec.

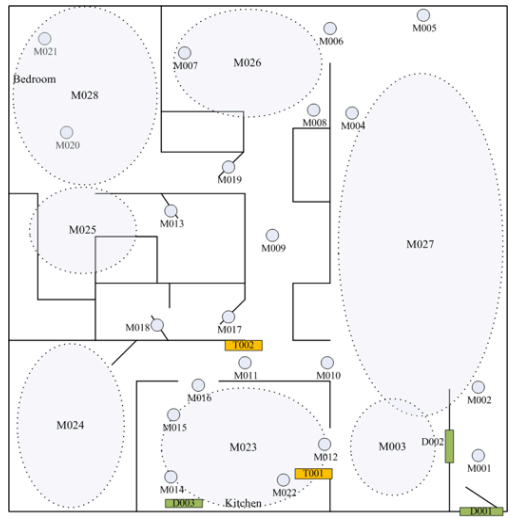

**Fig. 3.** The layout and sensor deployment of smart home Milan [32]. Note that the sensing coverage of an area motion sensor is represented by an oval. The sensing coverage of the other sensors is individually represented by a circle.

Fig. 4(a) shows the temporary global sensor topology deduced by PMA. Note that a confidence value is presented next to each directed edge. A higher confidence value from sensor $A$ to sensor $B$ implies that PMA has more confidence that an elder is able to directly move from sensor $A$ to sensor $B$, where $A$ and $B$ are any two different sensors in Milan. Fig. 4(b) illustrates the final global sensor topology deduced by PMA. Note that threshold $\alpha = 13 \ (= \left\lfloor \frac{5597}{415} \right\rfloor)$. Compared with Fig. 4(a), we can see that some edges disappear because their confidence values satisfy neither Rule 1 nor Rule 2.

From Fig. 4(b), we observe that PMA made some false deductions. For instances, there should be a bidirectional edge between M005 and M006 because the woman living in Milan should be able to directly move from M005 to M006 and vice versa according to the layout of Milan (see the upper right corner of Fig. 3). However, PMA was unable to deduce this relationship. Besides, we can see that PMA was unable to deduce any relationship between M024 and the rest of the sensors. The main reason is that the confidence values between M024 and all other sensors are all lower than the predefined threshold (see Fig. 4(a)), implying that the woman did not have much movement to the place where M024 is deployed during the observation period.

(a) The temporary global sensor topology

(b) The final global sensor topology

**Fig. 4.** The global sensor topology of Milan.

**Fig. 5.** The sensor spatial relationships of Milan and those reasoned by PMA. Note that all false deductions are highlighted.

Fig. 5 shows the accuracy of PMA on reasoning about the sensor relationships of Milan where
1. 1/1 in attribute (*A/B*) means that sensor *B* is directly reachable from sensor *A* according to the layout of Milan, and PMA is able to deduce a direct edge from *A* to *B* (i.e., $A \rightarrow B$).
2. 1/0 in attribute (*A/B*) means that *B* is directly reachable from *A* based on the layout, but PMA is unable to deduce direct edge $A \rightarrow B$.
3. 0/0 means that *B* is unreachable directly from *A* based on the layout, and PMA is also unable to deduce direct edge $A \rightarrow B$.
4. 0/1 means that *B* is unreachable directly from *A* based on the layout, but PMA is able to deduce direct edge $A \rightarrow B$.

It is clear that PMA makes an incorrect deduction when 1/0 or 0/1 appears. Recall that the total number of sensors in Milan is 31, so the total number of sensor-relationship deductions is 930 (=31*31-31). In which, the total number of false deductions is 23 (see all the highlights in Fig. 5), meaning that the accuracy rate of PMA on deducing sensor-relationship is 97.5% $\cong \left(1 - \frac{23}{930}\right) * 100\%$.

PMA used the algorithm shown in Fig. 2 to mine sensors in the bedrooms and kitchen/dining room of Milan. PMA first extracted all indoor activities between 2 am and 6 am from each of the six days. The total number of such activities is four, and the corresponding sensor-ID lists are shown in Fig. 6. After applying ARL with minSupport of 0.5 on the four sensor-ID lists, PMA returned a sensor set {M013, M020, M021, M025, M028}. This is because 1) the occurrence ratio of this set in the four lists (i.e., 2/4) satisfies the minSupport, 2) this set is the largest one, and 3) this set has the highest number of occurrences as compared with other sets with the same size. Therefore, {M013, M020, M021, M025, M028} was reasoned to be sensors in a bedroom. Since all other sensors in the topology except M019 do not have bidirectional edges with a half of this sensor set, M019 was also reasoned to be a sensor in the same bedroom. In other words, {M013, M019, M020, M021, M025, M028} are all the sensors that PMA found in the bedroom.

| | |
|---|---|
| 1 | M013 M020 M021 M025 M028 |
| 2 | M013 M020 M021 M025 M028 |
| 3 | M020 M021 M025 M028 |
| 4 | M020 M025 M028 |

**Fig. 6.** All sensor-ID lists between 2 am and 6 am of the six days for deducing bedroom sensors in Milan.

After that, PMA attempted to mine sensors in another bedroom by discarding any sensor-ID list that contains any known bedroom sensors (i.e., M013, M019, M020, M021, M025, and M028). The resulting sensor-ID list was empty, implying that there is no other bedroom in Milan. As compared with the sensor deployment shown in Fig. 3, we can see that PMA made a correct reasoning, i.e., there is only one bedroom in Milan, and M013, M019, M020, M021, M025, and M028 are indeed deployed in the bedroom.

PMA continues to extract all indoor activities between 6 pm and 7 pm from each of the six days to mine kitchen/dining room sensors. Fig. 7 lists all the corresponding sensor-ID lists. After applying ARL with minSupport of 0.5 on these lists, PMA returned a matching sensor set, i.e., {M014, M015, M022, M023}. All sensors in this set were therefore reasoned as being sensors deployed in the kitchen/dining room of Milan. Furthermore, since only D003, M012, and M016 in the global topology have bidirectional edges with a half of {M014, M015, M022, M023}, these three sensors were all deduced as sensors in the kitchen/dining room as well. By comparing the above result with the layout shown in Fig. 3, this reasoning is confirmed correct.

Based on all the above results, we demonstrate that PMA is able to mine a lot of privacy about Milan, including the global sensor topology, sensors in the bedroom, and sensors in the kitchen/dining room. This information might be mistreated by malicious people and might therefore potentially harm elders' safety.

```
1   M012 M014 M015 M022 M023
2   M014 M015 M022 M023
3   M008 M009 M011 M014 M015 M016 M017 M022 M023 M026
4   M005 M006 M008 M009 M011 M013 M019 M025 M026
5   M011 M014 M015 M016 M017 M018 M022 M023
6   M003 M005 M007 M008 M009 M011 M012 M014 M015 M016 M023 M026
7   M009 M011 M014 M015 M016 M022 M023
8   M007 M008 M009 M011 M014 M015 M016 M019 M022 M023 M026
9   M008 M009 M014 M015 M016 M017 M022
10  M014 M015 M022 M023
11  M014 M015 M022 M023
12  M014 M015 M023
13  M003 M004 M012 M014 M015 M022 M023 M027
14  D003 M012 M014 M015 M022 M023
15  M005 M006 M007 M008 M026 M027
16  M012 M014 M022 M023
17  M014 M022 M023
18  M001 M002 M003 M012 M014 M023
19  D003 M012 M014 M023
20  M001 M002 M003 M012 M014 M015 M022 M023 M027
21  M002 M003 M004 M027
22  M012 M014 M015 M022 M023
23  D003 M012 M014 M015 M022 M023
24  M014 M015 M023
25  M014 M015 M023
26  M014 M015 M022 M023
27  M012 M014 M015 M022 M023
28  M014 M015 M022 M023
29  D003 M011 M014 M015 M016 M017 M022 M023
30  M001 M014 M015 M022 M023
31  M001 M002 M003 M012 M014 M022 M023 M027
32  M001 M002 M003 M006 M007 M008 M009 M010 M012 M014 M022 M023 M026 M027
33  M012 M014 M015 M022 M023
34  M003 M012 M014 M015 M022 M023
35  M014 M022 M023
36  M012 M014 M015 M022 M023
37  M014 M015 M022 M023
38  D003 M014 M015 M022 M023
39  M014 M022 M023
40  M001 M002 M003 M014 M015 M022 M023
41  M012 M022 M023
```

**Fig. 7.** All sensor-ID lists between 6 pm and 7 pm of the six days for deducing kitchen/dining room sensors in Milan.

## 5    Conclusions and Future Work

In this paper, we have proposed PMA for mining global sensor topology and sensor locations from a smart home. Our experimental results demonstrate that PMA is capable of mining a global sensor topology of a smart home and reason about most bedroom sensors and kitchen/dining room sensors. The key factor is that PMA successfully defines rules for identifying indoor activities and constructing the spatial relationships between different sensors. The deduction logic of PMA is based on two reasonable assumptions, namely that people usually stay in their bedrooms between 2 am and 6 am, and that people usually stay in their kitchens/dining rooms between 6 pm and 7 pm. This is another key to success. The experimental results show that PMA is capable of mining lots of private information from a smart home using very limited information.

In the future, we would like to further extend PMA by reasoning about other locations, such as entrances or toilets. We are also interested in deducing elders' daily pattern, e.g., how often they stay in their bedrooms and kitchens/dining rooms, when elders go out and come home, etc. Furthermore, we would like to find out the best values for the parameters $x$ and $y$.

**Acknowledgments.** This work is supported by the project IoTSec – Security in IoT for Smart Grids, with number 248113/O70 part of the IKTPLUSS program funded by the Norwegian Research Council. This work is also partially supported by SIRIUS, which is a Norwegian Centre for Research-driven Innovation in Norway. The authors would like to thank the anonymous reviewers for their valuable comments and suggestions that improve the paper quality.